\documentclass[aps,prb,showpacs,groupedaddress,twocolumn]{revtex4}
\usepackage{amsmath,amssymb}
\usepackage{graphicx} 
\usepackage{dcolumn} 
\usepackage{bm} 

\begin{document}
\draft
\title{\bf Response time of a normal-superconductor hybrid
system under the step-like pulse bias}

\author{Yanxia Xing$^1$, Qing-feng Sun$^{1,\ast}$,
and Jian Wang$^{2}$}

\address{
$^1$Beijing National Lab for Condensed Matter Physics and
Institute of Physics, Chinese Academy of Sciences, Beijing 100080,
China\\
$^2$Department of Physics and the Center of Theoretical and
Computational Physics, The University of Hong Kong, Pokfulam Road,
Hong Kong, China}

\begin{abstract}

The response of a quantum dot coupled with one normal lead and a
superconductor lead driven by a step-like pulse bias $V_L$ is studied
using the non-equilibrium Green function method. In the linear
pulse bias regime, the responses of the upwards and downwards bias
are symmetric. In this regime the turn-on time and
turn-off time are much slower than that of the normal system due to
the Andreev reflection. On the other hand, for the large pulse
bias $V_L$, the instantaneous current exhibits oscillatory
behaviors with the frequency $\hbar\Omega =qV_L$. The turn
on/off times are in (or shorter than) the scale of $1/V_L$, so
they are faster for the larger bias $V_L$. In addition, the
responses for the upwards and downwards bias are asymmetric at
large $V_L$. The turn-on time is larger than the turn-off time
but the relaxation time \cite{note1} depends only on the coupling
strength $\Gamma$ and it is much smaller than the turn-on/off times
for the large bias $V_L$.

\end{abstract}

\maketitle

\section{introduction}

In the past two decades, nanoscopic physics has developed significantly
and becomes an active field of condensed-matter physics. The
quantum transport property also becomes one of the most interesting
phenomena in nanoscopic physics because of the possibility of
designing and fabricating artificial setups in the nanometer
scale. Based on the transport physics in nanoscopic system, a
rich field for basic and applied research is
opened.\cite{book1} Furthermore, the time dependent nanoscopic
transport, in which the external time dependent fields drives the
electrons tunnel through a nanoscopic system, has received increasing
attention in recent years. The main feature of the transport in the nanometer
scale is that the electron keeps the phase coherence when traversing
through the device. While the external time
dependent field affects the phase factor of the incident electron
differently in different parts of the system.\cite{Jauho2}
If the external time dependent field is sinusoidal (e.g. microwave radiation),
an electron can tunnel through the system by emitting or absorbing photons
giving rise to the photon-assisted tunnelling (PAT). Electron transport with PAT
has been extensively investigated for various systems, such as single or two
coupled quantum dot (QD),\cite{Kouwenhoven,Sun1,Oosterkamp} Kondo
regime,\cite{Ng} hybrid system,\cite{Sun2} and so on. For transient transport,
one of the most interesting issues is how fast can a device turn on or
turn off a current. With the development of the molecular devices, there is
clearly a need to technologically provide a particular viable switching device.
Indeed, some recent experimental and theoretical works have
already begun to study the response of ac signals of the molecular
devices.\cite{Burke} Consequently, a step or pulsed ac signals
are the simplest choice, since it can provide a less ambiguous
measure of time scales. For this reason, the pulsed field was
studied in a variety of systems, including Kondo regime,
\cite{Plihal,Schiller} a single QD,\cite{Jauho1} or nano
structure,\cite{Maciejko,Zhu}.

So far, the study of response of pulsed bias is only focused on normal
nanostructures. Since the interplay between
nanoscopic physics and the physics of superconductivity has made the hybrid
structure a very fruitful research field,\cite{Lambert} it will
be interesting to study the dynamic response a hybrid structure with a
superconductor lead where the Andreev reflection is present near the
normal-superconductor (N-S) interface. Indeed, there are many
interesting phenomena in the N-S hybrid systems. First of all,
because there exists an energy gap $\Delta$ in the
superconductor, an incident electron from the normal side with
energy $\epsilon$ inside the gap $\Delta$ can not tunnel into
the superconductor. But the tunneling can occur via a two-particle process,
in which the incident electron is reflected as a hole with the energy
$-\epsilon$. At the same time, a Cooper pair is created in the
superconductor region. This is the Andreev
reflection.\cite{Andreev} Secondly, for the superconductor-normal
region-superconductor (S-N-S) system, Andreev bound states form in
normal region due to the Andreev reflections at N-S interfaces.\cite{Kulic}
These bound states exist in pairs, and a Josephson supercurrent can flow
through the S-N-S system which is carried by the Andreev bound state.\cite{Sols}
Thirdly, when the S-N-S device is under an
external dc bias $V$, an ac current with frequency $\omega=2|e|V$
appears. The time-average current versus bias $V$ exhibit the
subharmonic gap structure when $eV < 2\Delta$.\cite{Kleinsasser}

In this paper, we explore the effect of Andreev reflection on the ac response
of hybrid system. Specifically, we investigate ac response of a quantum dot (QD)
with a single level $\epsilon_0$ connected by a normal and a superconductor
lead (N-QD-S). For simplicity, we consider a large QD so that the intra-dot
electron-electron (e-e) is weak and can be neglected.\cite{Kondo}
The transient transport is driven by a pulsed bias potential $W(t)$.
For simplicity, the ac pulsed bias is only added in the left lead,
and we set $W_R(t)=0$. We consider two different pulsed bias: (i)
upwards pulse with $W_L(t)=0$ for $t<0$ and $W_L(t)=V_L$ otherwise.
(ii) downwards pulse with $W_L(t)=V_L$ when $t<0$ and $W_L(t)=0$
otherwise. For normal structures, Wingreen {\sl et al.} presented a general
formula for the current driven by the time dependent external fields
by using the non-equilibrium Green function (NEGF)
method.\cite{Jauho2,Jauho1} With this general formula the
time dependent current driven by the ac pulse
can be calculated. For hybrid structures, the system is in steady
state at $t<0$ and the current is time independent. At $t=0$, bias
is abruptly turned on for the upwards pulse case or turned off for
the downwards pulse case. After that, the system begins to relax and
the Andreev reflection plays an important role in the relaxation
process. Finally, the system enters into a new steady state. We find
that, the relaxation time depends on the coupling strength and
is slower in the N-QD-S system (named hybrid system hereafter) than
in the N-QD-N system (named normal system hereafter). In the linear bias
regime, the rising and falling processes are symmetric so that the turn-on
time is same as the turn-off time. In this regime, the Andreev reflection
is important. As a result, the instantaneous current shows a clear increase
(decrease) before reaching the new steady state for the downwards (upwards) pulse.
For the large bias case, the time dependent current oscillates
with the frequency $\omega=qV_L$. In this regime, the upwards and the
downwards processes are asymmetric and the turn-on time is much
larger than the turn-off time. In this nonlinear regime, the Andreev
process is negligible and the current in the hybrid system
is close to that of the normal system.

The rest of this paper is organized as follows: In Sec.II, the
theoretical formula for calculating the time dependent current in
N-QD-S system is presented. To understand the numerical results, the
current away from the current at $t=0$ is
expanded to the first order in the external bias. In
Sec.III, we show the numerical results along with some discussions.
Finally, the brief summary is given in Sec. IV.

\section{theoretical formula}
Considering a hybrid system that consists of a QD
coupled to a normal metal lead and a superconductor lead with the
external time dependent bias potential $W_L(t)$ that is added only on
the left normal lead. The Hamiltonian of the system is written as
follows:
\begin{equation}
H=H_L+H_R+H_{D}+H_T
\label{total}
\end{equation}
where $H_L$ and $H_R$ describe the left normal lead and the
right superconductor lead, respectively. $H_{D}$ is
Hamiltonian of the isolated central QD, and the $H_T$ couples the
left and right leads to the QD. They can be written in the
following forms:\cite{Cuevas,Sun}
\begin{eqnarray}
H_L&=&\sum\limits_{k \sigma} (\epsilon_{L,k}+W_L(t)) C^{\dagger}_{L,k \sigma}
C_{L,k \sigma} \nonumber \\
H_R &=& \sum\limits_{k \sigma} \epsilon_{R,k} C^{\dagger}_{R,k \sigma}
C_{R,k \sigma} + \nonumber \\
& & \sum\limits_k [\Delta C_{R,k \downarrow} C_{R,-k \uparrow} +
\Delta C^{\dagger}_{R,-k \uparrow} C^{\dagger}_{R,k \downarrow}]
\nonumber\\
H_{D}& = &\sum\limits_{\sigma}
\epsilon_0 d^{\dagger}_{\sigma} d_{\sigma}
\nonumber\\
H_T &=& \sum\limits_{\sigma,k,\alpha} t_{k,\alpha}
C^{\dagger}_{\alpha,k \sigma}d_{\sigma}+h.c.,
\label{H}
\end{eqnarray}
where $\alpha=L,R$. The operator $d_\sigma$ and $C_{\alpha,k
\sigma}$ destroy an electron with spin $\sigma$ in the QD and in
the left or right lead, respectively. For simplicity, we only
consider a single level in the QD and neglect intradot
electron-electron Coulomb interaction. Under the adiabatic
approximation, the time-dependent bias potential can be included
in the single electron energy $\epsilon_{L,k}(t)$. We separate
$\epsilon_{L,k}(t)$ into two parts: $\epsilon_{L,k}$ and $W_L(t)$,
where $\epsilon_{L,k}$ is the time-independent single electron
energy and $W_L(t)$ is a time dependent part from the external
time dependent bias potential. In this paper, $W_L(t)$ is the
step-like pulse with two different forms: (i) upwards pulse with
$W_L(t)=0$ when $t<0$ and $W_L(t)=V_L$ otherwise, (ii)
downwards pulse with $W_L(t)=V_L$ when $t<0$ and $W_L(t)=0$
otherwise. These two types of pulse describe the system abruptly
turned on or turned off at time $t=0$. $\Delta$ in the Hamiltonian
$H_R$ is the superconducting energy gap. We assume that $\Delta$ is
a real parameter by selecting a special phase of the
superconductor lead in our calculation.\cite{book} Due to the
existence of the superconducting lead, it is convenient to introduce the
Nambu representation.\cite{Nambu} In the Nambu representation, the
Fermi energy of the left normal lead is set at the superconducting
condensate and for the spin down electron the energy is negative
and is viewed as the hole. So, the Hamiltonian in Eqs.(\ref{H})
can be rewritten in the matrix form:
\begin{eqnarray}
&&H_L = \sum_k \Psi^\dagger_{L,k}
\left(
\begin{array} {cc}
\epsilon_{L,k}+W_L(t) & 0  \\
0 & -\epsilon_{L,-k} -W_L(t)
\end{array}
\right)
\Psi_{L,k}
\nonumber \\
&&H_R = \sum_k \Psi^\dagger_{R,k}
\left(
\begin{array} {cc}
\epsilon_{R,k} & \Delta   \\
\Delta & -\epsilon_{R,-k}
\end{array}
\right)
\Psi_{R,k}
\nonumber \\
&&H_{D} =  \Phi^\dagger
\left(
\begin{array} {cc}
\epsilon_0 &~~ 0  \\
0 &~~ -\epsilon_0
\end{array}
\right)
\Phi
\nonumber \\
&&H_T = \sum_{k,\alpha} \Psi_{k,\alpha}^\dagger
\left(
\begin{array} {cc}
t_{k,\alpha,\uparrow} &~~ 0  \\
0 &~~ -t_{k,\alpha,\uparrow}^*
\end{array}
\right)
\Phi
+ H.C.,
\end{eqnarray}
where
\begin{eqnarray}
\Psi_{\alpha,k}=
\left(
\begin{array} {l}
C_{\alpha,k \uparrow} \\
C^\dagger_{\alpha,-k \downarrow}
\end{array}
\right),~~~
\Phi=\left(
\begin{array} {c}
d_\uparrow \\
d^\dagger_{\downarrow}
\end{array}
\right).
\label{define}
\end{eqnarray}

The current from the left lead to the QD can be calculated from
the evolution of the number operator of the electrons in the left
lead, $N_{{L,\uparrow(\downarrow)}}=\sum_k C^\dagger_{\alpha,k
\uparrow(\downarrow)} C_{\alpha,k
\uparrow(\downarrow)}$.\cite{Sun,Jauho1,Jauho2} Using the Keldysh
equation and the theorem of analytic continuation, the current
through the left normal metal lead is expressed
as:\cite{Jauho2,Sun,Xing}
\begin{eqnarray}
J_L(t)&=&-2q {\bf Re}\int^t_{-\infty} dt'
   \{[G^r(t,t')\Sigma^<_L(t',t)\nonumber \\
   &&+G^<(t,t')\Sigma^a_L(t',t)]_{11}-[G^r(t,t')\Sigma^<_L(t',t)\nonumber \\
&&+G^<(t,t')\Sigma^a_L(t',t)]_{22}\}
\label{Cur0}
\end{eqnarray}
Here the Green function $G^{r/<}$ and the self-energy $\Sigma^{</a}$
are all two dimensional matrices in the Nambu representation.
Since the spin up and spin down are symmetric in the
Hamiltonian, the current contributed by the electrons with
spin up is same as the current by the spin down electrons.
Consequently, the current is given by:
\begin{eqnarray}
J_L(t)&=&-4q {\bf Re}\int^t_{-\infty} dt'
   [G^r(t,t')\Sigma^<_L(t',t)\nonumber\\
   &&+G^<(t,t')\Sigma^a_L(t',t)]_{11}.
   \label{Cur1}
\end{eqnarray}
Because of
$\Sigma^a_L(t,t')=[\Sigma^r_L(t,t')]^\dagger=(i \Gamma_L/2)\delta(t-t'){\bf I} $
(see Appendix) where ${\bf I}$ is the $2\time 2$ unit matrix. Note that
only $G^<(t,t)$ instead of $G^<(t,t')$ is needed in the
Eq.(\ref{Cur1}). By using the Keldysh equation $G^<=G^r\Sigma^<G^a$
with the self-energies obtained in the appendix, the Green
function $G^<(t,t)$ can be solved:
\begin{eqnarray}
G^<(t,t) &= &\sum_{\alpha}\int dt_1 \int dt_2 G^r(t,t_1)
\Sigma_\alpha^<(t_1,t_2)G^a(t_2,t)\nonumber \\
&= &i \int \frac{d \omega}{2\pi}f(\omega)
G^r(\omega){\bar \Gamma}_R(\omega)G^a(\omega)+ \nonumber \\
&& i\sum_{\sigma} \int \frac{d \omega}{2\pi}f(\omega)
A_{L,\sigma}(\omega,t)s_\sigma
\Gamma_L(\omega)A^+_{L,\sigma}(\omega,t). \nonumber \\
\label{Gl}
\end{eqnarray}
where $\sigma=\pm1$ denotes the spin up $\uparrow$ and spin down
$\downarrow$,
$$ {\bar \Gamma}_R(\omega)=\theta(\omega-\Delta)
\frac{\Gamma_R}{\sqrt{\omega^2-\Delta^2}}\left(
\begin{array}{cc}
|\omega| & \Delta \\
\Delta & |\omega|
\end{array}
\right),$$
and
\begin{equation}
 s_{\uparrow}=\left( \begin{array}{cc}
1 & 0 \\
0 & 0
\end{array}
\right),~~~~
 s_{\downarrow}=\left(
\begin{array}{cc}
0 & 0 \\
0 & 1
\end{array}
\right),
\end{equation}
\begin{eqnarray}
A_{L,\sigma}(\omega,t)= \int^t_{-\infty} dt_1
G^r(t,t_1)e^{i\omega(t-t_1)+i\sigma\int_{t_1}^t dt_2 W_L(t_2)},
\label{A}
\end{eqnarray}
The Green functions $G^{r/a}(\omega)$ in Eq.(\ref{Gl}) are the
Fourier transformation of $G^{r/a}(t,t')$ with
$G^{r/a}(\omega)=\int d(t-t') e^{i\omega(t-t') }G^{r/a}(t,t')$.
Notice that in the present system the retarded and advanced Green
functions $G^{r/a}(t,t')$ are still the function of the time
difference $t-t'$, although there exists the time dependent bias
$W_L(t)$, since $G^r(\omega)$ can be obtained from Dyson equation:
\begin{eqnarray}
G^r(\omega)&=&[\omega-H_{dot}-\Sigma_L^r-\Sigma_R^r]^{-1}\nonumber \\
&=&\frac{1}{Det}\times \left(
\begin{array}{cc}
B_{11} & i\nu \Gamma_R \beta'/2 \\
i\nu \Gamma_R \beta'/2 & B_{22}
\end{array}
\right) \label{Grw}
\end{eqnarray}
where $B_{11}=\omega+\epsilon_0+i\Gamma_L/2+i\nu \Gamma_R
\beta/2$, $B_{22}=\omega-\epsilon_0+i\Gamma_L/2+i\nu \Gamma_R
\beta/2$, $\beta=\Delta/\sqrt{\omega^2-\Delta^2}$,
$\beta'=\omega/\sqrt{\omega^2-\Delta^2}$,
$Det=B_{11}B_{22}+(\Gamma_R \beta')^2/4$, and $\nu=1$ for
$\omega>-\Delta$ and $\nu=-1$ otherwise. In the above derivation,
the wide-band limit has been used and $\Gamma_{\alpha}$ are
assumed independent of $\omega$.\cite{foot11} It also is worth
mentioning that the Green function $G^{r/a}(\omega)$ is not
affected by the time-dependent bias potential $W_L(t)$.

Substituting $G^<(t,t)$ [in Eq.(\ref{Gl})] and the self-energies
$\Sigma^{</a}(t',t)$ (in appendix) into Eq.(\ref{Cur1}), the
time-dependent current $J_L(t)$ is obtained straightforwardly.
Similar to the work in the normal system by Wingreen, Jauho, and
Meir,\cite{Jauho2} the current $J_L(t)$ can also be split into two
terms $J_L^{in}(t)$ and $J_L^{out}(t)$:
\begin{eqnarray}
&J^{in}_L(t)=4q\int \frac{d \omega}{2\pi}f(\omega){\bf Im}
\{\Gamma_L [A_{L \uparrow}(\omega,t)]_{11}\}
~~~~~~~~~~~&\nonumber \\
&J^{out}_L(t)=-2q \int \frac{d \omega}{2\pi} f(\omega)
{\bf Re}\{\Gamma_L[G^r(\omega){\bar \Gamma}_R(\omega)G^a(\omega)+
&\nonumber \\
&~~~ \sum_\sigma A_{L \sigma}(\omega,t)s_\sigma \Gamma_L A^+_{L
\sigma}(\omega,t)]_{11}\}. &\label{Cur2}
\end{eqnarray}
and $J_L(t)=J_L^{in}(t)-J_L^{out}(t)$. Here the current
$J_L^{in}(t)$ is contributed by the electrons tunnelling from the
left lead to the empty QD, and the current $J_L^{out}(t)$
describes the electrons tunnelling from the QD to the empty left
lead, so they have the opposite sign.\cite{Jauho2}

The above formulations [Eqs.(\ref{A},\ref{Grw},\ref{Cur2})] for
calculating the current are valid for any time-dependent bias
$W_L(t)$. In the following, two special cases for upwards and downwards
pulses $W_L(t)$ are substituted into these formulations to
obtain $A_{L\sigma}(\epsilon,t)$ [Eq.(\ref{A})] and then the
currents $J_{L}^{in}(t)$ and $J_{L}^{out}(t)$ [Eqs.(\ref{Cur2})].

For the downwards pulse with $W_L(t<0)=V_L$ and $W_L(t>0)=0$,
$A_{L\uparrow}(\epsilon,t)$ is found to be:
\begin{eqnarray}
A_{LD,\uparrow}(\omega,t<0) & = &G^r(\omega+V_L)\nonumber \\
A_{LD,\uparrow}(\omega,t>0)& = &G^r(\omega)+\int \frac{dE}{2\pi
i}e^{-i(E-\omega)t}G^r(E)
\nonumber \\
 && \left[\frac{1}{E-\omega-V_L-i0^+}-\frac{1}{E-\omega-i0^+}\right].
\nonumber \\ \label{A11}
\end{eqnarray}

For the upward pulse with $W_L(t<0)=0$ and $W_L(t>0)=V_L$,
$A_{L\uparrow}(\epsilon,t)$ is:
\begin{eqnarray}
A_{LU,\uparrow}(\omega,t<0)& = &G^r(\omega)
\nonumber \\
A_{LU,\uparrow}(\omega,t>0) &= &G^r(\omega+V_L)-\int
\frac{dE}{2\pi i} e^{-i(E-\omega-V_L)t}G^r(E)
\nonumber \\
&&
\left[\frac{1}{E-\omega-V_L-i0^+}-\frac{1}{E-\omega-i0^+}\right].
\nonumber \\ \label{A22}
\end{eqnarray}
Here $A_{L\sigma}(\epsilon,t)$ for the downward and upward pulse
biases have been labelled by $A_{LD,\sigma}(\epsilon,t)$ and
$A_{LU,\sigma}(\epsilon,t)$, respectively. For $t<0$, the system
is in the steady state, so $A_{LD,\sigma}(\epsilon,t)$ and
$A_{LU,\sigma}(\epsilon,t)$ are independent of time $t$. On
the other hand, for $t>0$, they are obviously dependent on
time $t$. For the purpose of numerical calculation, we rewrite
$A_{LD/U,\sigma}(\epsilon,t)$ for $t>0$ in the following form by
using the residue theorem:
\begin{eqnarray}
&&A_{LD,\uparrow}(\omega,t>0) = G^r(\omega)+ \nonumber \\
&&e^{-iV_L t}\int_t^\infty d\tau
e^{i(\omega+V_L)\tau}G^r(\tau) -\int_t^\infty d\tau
e^{i\omega \tau}G^r(\tau),
\nonumber \\ 
&& A_{LU,\uparrow}(\omega,t>0) = G^r(\omega+V_L)+\nonumber \\
&& e^{iV_L t}\int_t^\infty d\tau
e^{i\omega\tau}G^r(\tau) -\int_t^\infty d\tau
e^{i(\omega+V_L)\tau}G^r(\tau). \nonumber \\ \label{A22}
\end{eqnarray}

The expressions of $A_{L,\downarrow}(\omega,t)$ are similar to
that of $A_{L\uparrow}(\omega,t)$ and can be obtained from
Eq.(\ref{A22}) by changing $V_L$ to $-V_L$. After
solving $G^r(\omega)$ and $A_{L\sigma}(\omega,t)$, the currents
$J_{L}^{in}(t)$ and $J_{L}^{out}(t)$ [Eq.(\ref{Cur2})] can be
calculated straightforwardly. In the limits $t \leq 0$ and $t\rightarrow
\infty$, the system is in the steady state.
$A_{LD,\sigma}(\omega,t)$ and $A_{LU,\sigma}(\omega,t)$ in
Eq.(\ref{A22}) then reduce to the value of the steady state in these
two limits and so is the current $J_L(t)$. For example, for the
downward pulse, $A_{LD,\sigma}(\omega,t)=G^r(\omega+\sigma
V_L)$ for $t\rightarrow 0$ , and
$A_{LD,\sigma}(\omega,t)=G^r(\omega)$ when $t\rightarrow \infty$.
Furthermore, the current $J_L(t)$ reduces to the one of the steady
case with dc bias $V_L$ when $t\leq0$ , and is zero when
$t\rightarrow \infty$. On the other hand, for the upwards pulse,
the current $J_L(t)$ is zero when $t\leq0$, and is same with the
steady state current with the dc bias $V_L$ in $t\rightarrow
\infty$ limit.

In the small pulse bias $V_L$ limits, we can expand
$A_{L\sigma}(\omega,t>0)$ to the first order of $V_L$ as:
$A_{L\sigma}(\omega,t>0) = A_{L\sigma}(\omega,t=0) +
A^1_{L\sigma}(\omega,t>0) V_L$. $A^1_{L\sigma}(\omega,t>0)$ can be
expressed as:
\begin{eqnarray}
&A^1_{LD,\sigma}(\omega,t>0)=& \nonumber \\
&-i\sigma t\int_t^\infty d\tau
e^{i\omega\tau}G^r(\tau)-\sigma \int_0^t d\tau  i\tau
e^{i\omega\tau}G^r(\tau) & \nonumber \\
&A^1_{LU,\sigma}(\omega,t>0)=& \nonumber \\
& i\sigma t\int_t^\infty d\tau
e^{i\omega\tau}G^r(\tau)+\sigma
\int_0^t d\tau  i\tau
e^{i\omega\tau}G^r(\tau)&
\label{ALDU}
\end{eqnarray}
From Eq.(\ref{ALDU}), we can see that
$A^1_{LD,\sigma}(\omega,t)=-A^1_{LU,\sigma}(\omega,t)$. This
means that the upwards pulse and
downwards pulse induce the same relaxation process in the small
pulse bias $V_L$ limits, except that the currents deduced from
them are relaxed in the opposite direction. Finally, the currents $J_L^{in}(t)$
and $J_L^{out}(t)$ in small $V_L$ limits can also be expanded as:
$J_L^{in/out}(t)=J_L^{in/out}(0)+X^{in/out}(t)V_L$. Here
$X^{in/out}(t)$ is the first order expansion coefficient with the
respect to $V_L$, and $X^{in/out}(t)$ is expressed as:
\begin{eqnarray}
&&X^{in}(t) = 4q \int \frac{d \omega}{2\pi}{\bf Im}
f(\omega)\Gamma_L\{A^1_{L,\uparrow}(\omega,t)\}_{11}\nonumber \\
&&X^{out}(t) = -2q \int \frac{d \omega}{2\pi}{\bf Re}
f(\omega)\Gamma_L \sum_\sigma \nonumber \\
          &&\{A^1_{L,\sigma}(\omega,t)s_\sigma \Gamma_L
              G^a(\omega)+
            G^r(\omega)s_\sigma \Gamma_L
             [A^1_{L,\sigma}(\omega,t)]^\dagger\}_{11}
             \nonumber \\
\end{eqnarray}

\section{numerical results and discussions}
In the numerical calculation, we set temperature to zero. In fact,
finite temperature only makes the current curve more smooth and
does not affect main features. We focus on the weak coupling case with
$\Gamma_{L/R} \ll \Delta$ and set $\Gamma=\Gamma_L+\Gamma_R=1$
as energy unit. The energy gap of the superconductor is
$\Delta=15$. The energy level $\epsilon_0$ in the central region
is assumed to be zero which is same to the right Fermi level.
Because at $t\leq 0$ the system is in the steady state and the
current is time independent, so we only plot the current $J_L(t)$
and the related quantities for $t \geq 0$ in the following discussion.

\begin{figure}
\includegraphics[bb=9mm 11mm 210mm 200mm, width=6.0cm,totalheight=5cm, clip=]{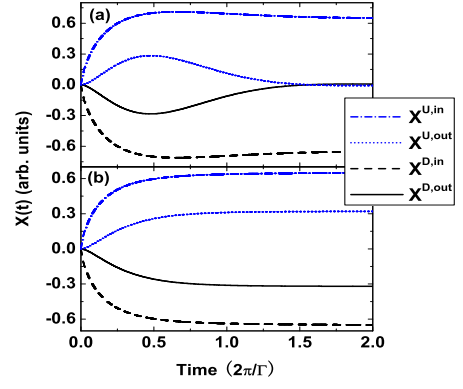}
\caption{(Color on line) The first order expansion coefficient
$X(t)$ of the current $J_L(t)-J_L(0)$ {\sl vs.} the time $t$ for
the downwards and upwards pulse bias case in the hybrid N-QD-S
system (a) and the normal N-QD-N system (b). The parameters are:
$\Gamma=1$, $\delta \Gamma=0$, $\Delta=15$, $\epsilon_0=0$.}
\end{figure}

First of all, we study the small pulse bias $V_L$ limit, in which the
instantaneous current $J_L(t)$ can be expanded as:
$J_L(t)=J_L(0)+X(t)V_L$, and we also take the symmetric barriers,
i.e., $\delta\Gamma=\Gamma_L-\Gamma_R=0$. The first-order
expansion parameters $X^{U/D,in/out}(t)$ of the currents
$J_L^{in}(t)$ and $J_L^{out}(t)$ versus the time $t$ are plotted
in Fig.1. Here the indices $U$ and $D$ denote the upwards and
downwards pulses, respectively. For comparison, we also show the
corresponding parameters $X^{U/D,in/out}(t)$ for the normal system
in Fig.1(b). From Fig.1, we can see that the expanding parameters
$X(t)$ for the upwards and downwards pulses are symmetric, i.e.
$X^{U,in/out}(t) = -X^{D,in/out}(t)$. It means that in the small
$V_L$ limit (i.e. the linear regime), the current turned off or
turned on by the downwards or upwards pulses in exactly the
same manner with the same time scale for both normal system and
hybrid system. In other words, the case of the downwards pulse is
the reversal process of the upwards pulse. So in the following, we
use the upwards pulse as an example in the linear region.

At time $t\le 0$, the driving bias is zero for the upwards case. The
system is in equilibrium state so the current $J_L^U$ is zero and
$J_L^{U,in}$ and $J_L^{U,out}$ cancel to each other. At $t=0$, the
bias is abruptly switched on. At $t>0$, the bias $W_L(t)$ is kept at
$V_L$ all along, the electrons with the energy in the bias window
begin to traverse through QD. As the time $t$ increases,
$J_L^{U,in}$ and $J_L^{U,out}$ deviate from the initial value
($t=0$). A net current gradually increases and the device is
gradually turned on. As a result, for the time $t$ from $0$ to about
$0.5(2\pi/\Gamma)$, $X^{U,in}(t)$ and $X^{U,out}(t)$ gradually
increase (see Fig.1). This increasing process is almost the same for
the normal system and the hybrid system. For the normal N-QD-N
device, the relaxation process completes near the time
$t=0.5(2\pi/\Gamma)$ and $X^{U,out}(t)$ is the half of $X^{U,in}(t)$
at large time. On the other hand, for the hybrid N-QD-S device, the
behavior of $X^{U,in}(t)$ is approximatively the same as that of
N-QD-N at large time, but $X^{U,out}(t)$ begins to decrease when
$t>0.5(2\pi/\Gamma)$, and it goes to zero at the end of the
relaxation process. So the current $J^U_L(t=\infty)$ for the N-QD-S
device is twice as large as that of the N-QD-N device. We interpret
these properties as follows. For the normal system, the fact that
$X^{in}(t=\infty)$ is twice of $X^{out}(t=\infty)$ is because
$X^{in}(t=\infty)$ and $X^{out}(t=\infty)$ are respectively
contributed by the electrons tunnelling from the left lead into the
empty QD and from the QD into the empty left lead with the
electronic energy $\omega$ between $0$ and $V_L$, and in this energy
range the distribution of the left lead is $f_L(\omega)=1$ but the
distribution in the QD is $(f_L(\omega)+f_R(\omega))/2 =1/2$ for
$t=\infty$. While for the hybrid N-QD-S system, after the bias is
turned on, the Andreev reflection begins to play a role. For
$J_L^{U,in}$, there is not much difference between the normal and
hybrid systems, since the electrons always tunnel from the left lead
into the QD in both systems. But for $J_L^{U,out}$, instead of
reflecting electrons from QD into the left lead in normal system,
the Andreev process reflects back the hole out of QD, which makes
$J_L^{U,out}$ decrease. Note that $T_A$ can be expressed as
\cite{Sun}:
\begin{equation}\label{Andr}
T_A=\frac{\Gamma^4}{64\omega^4+(\Gamma^2 + \delta\Gamma^2)^2},
\end{equation}
in the small bias limit ($\omega\approx0$) and $\delta\Gamma=0$,
nearly all of the incoming electrons participate in the Andreev
reflection. Because of this, $J_L^{U,out}(t=\infty)$ goes back to
the initial (t=0) value. So $X^{out}(t)$ decreases to zero at
$t=\infty$.

\begin{figure}
\includegraphics[bb=11mm 10mm 194mm 153mm, width=8cm,totalheight=5.6cm, clip=]{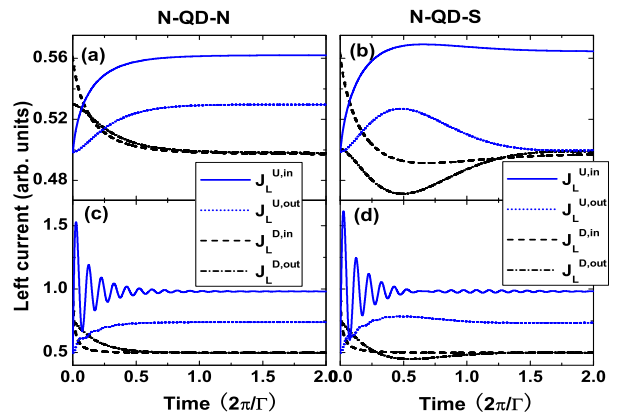}
\caption{(Color on line) The currents $J_L^{U/D,in}$ and
$J_L^{U/D,out}$ {\sl vs.} the time $t$ for the small pulse bias
$V_L=0.1$ (upper panels (a) and (b)) and the large pulse bias
$V_L=10$ (lower panels (c) and (d)) in upwards and downwards pulse
case. The left panels (a) and (c) are for the N-QD-N system and
the right panels (b) and (d) are for the N-QD-S system. The other
parameters are same with Fig.1}
\end{figure}

Next, we study the case of large pulse $V_L$. Fig.2(c) and (d)
depict the currents $J_L^{out}$ and $J_L^{in}$ versus time $t$ for
the large pulse strength $V_L=10$. For comparison, $J_L^{out}$ and
$J_L^{in}$ for the small pulse strength $V_L=0.1$ are also plotted
in the Fig.2(a) and (b). The currents $J_L^{out}$ and $J_L^{in}$
in the large bias case have the following characteristics: (i) In
the small bias limit, the relaxation processes of upwards and
downwards are symmetric. However, in the large pulsed bias $V_L$
case, they are asymmetric (see Fig.2c and 2d). For larger pulse
bias $V_L$, the asymmetry are stronger. (ii) For the large bias
case, $J_L^{U,in}$ for the upwards pulse oscillates with the
frequency $\hbar\Omega=qV_L$, which can be clearly seen in Fig.2c
and 2d for $V_L=10$. At $V_L=0.1$ the oscillation disappears
because $\hbar\Omega=qV_L$ is too small to oscillate before the system
is completely relaxed. (iii) $J_L^{U,out}$ ($J_L^{D,out}$) of
hybrid system increases (decreases) in the first and then
decreases (increases), and it reaches maximum (minimum) before the
current relaxed completely. This is different from the normal
system, in which the currents $J_L^{U,out}$ and $J_L^{D,out}$ are
monotonously relaxed into the steady state. (iv) The decreasing
(increasing) process of the current $J_L^{U,out}$ ($J_L^{D,out}$) in
the large bias case is much weaker than that of the small bias case
(see Fig.2b and d). Because for the large pulse, the
energy of the incident electrons $\omega$ is large, then $T_A \ll 1$
from Eq.(\ref{Andr}) and the Andreev reflection is weak. So most
of the incident electrons participate in the normal reflection.
Consequently, $J_L^{out}$ is humped up (or down) slightly.

\begin{figure}
\includegraphics[bb=10mm 11mm 199mm 203mm, width=6cm,totalheight=6cm, clip=]{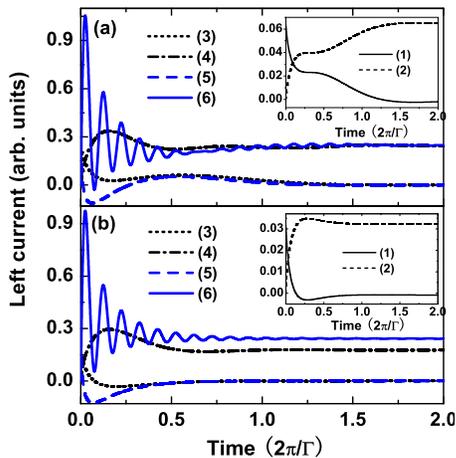}
\caption{(Color on line) The currents $J_L(t)$ {\sl vs.} the time
$t$ for the N-QD-S system (a) and the N-QD-N system (b) with the
different pulsed bias $V_L$. Main Figure is for the case of
$V_L=1$ and $V_L=10$. The case of $V_L=0.1$ is plotted in inset
panel. The curves are labelled as: (1) $J_L^D(V_L=0.1)$; (2)
$J_L^U(V_L=0.1)$; (3) $J_L^D(V_L=1.0)$; (4) $J_L^U(V_L=1.0)$; (5)
$J_L^D(V_L=10)$; (6) $J_L^U(V_L=10)$. The other parameters are
the same as Fig.1}
\end{figure}

Since the currents $J_L^{out}$ and $J_L^{in}$ can not be observed
independently, in the following we study the total current $J_L(t)$
($J_L=J_L^{in}-J_L^{out}$) which can be measured in the
experiment. Fig.3 shows the current $J_L^{U,D}$ driven by the
upwards and downwards pulses versus the time $t$ for the
different pulse strengths $V_L$. Here the current responses to the
upwards and the downwards pulse are symmetric at small
linear bias $V_L$ (see inset of Fig.3), but are asymmetric at
the large bias $V_L$ (see main of Fig.3). At the large $V_L$,
$J_L^{U}$ oscillates with the frequency $\hbar\Omega=V_L$. On the
other hand, $J_L^{D}$ always changes slowly regardless of the
large and small $V_L$.

Now we focus the turn on/off time (or rise/fall time
\cite{Plihal}) and the relaxation time (or saturation time
\cite{Schiller}). The former describes how fast can a device turn
on/off a current, which is necessary to provide a particular
viable switching device, and the latter was referred to how fast
can the device goes to a new steady state after a bias is abruptly
switched on. For the small bias $V_L$, the turn-on time, turn-off
time, and the relaxation time are almost same regardless of the
normal and hybrid systems. However these (turn on/off or
relaxation) times for the normal N-QD-N device are much shorter
than that of the hybrid N-QD-S device. For the normal device, it
has been well turned on or off at $t=0.2(2\pi/\Gamma)$. But for
the hybrid device, the system is turned on or off until $t=1.0
(2\pi/\Gamma)$. On the other hand, for the large bias, the
current $J_L(t)$ of the hybrid system has the same character with that
of the normal system, so do the turn-on/off time and the
relaxation time. Note that these three time scales are not equal now. The
turn-on time is the fastest, even faster than the scale $1/V_L
(2\pi/\Gamma)$. The turn-off time is in the scale $1/V_L
(2\pi/\Gamma)$, which is longer than the turn-on
time.\cite{Plihal} The relaxation is $\sim 0.5(2\pi/\Gamma)$,
which is the longest and only depends on the coupling strength
$\Gamma$. Let us explain why the character of $J_L(t)$ for the
normal and hybrid system are the same at large $V_L$ but very
different at small $V_L$. Because at the large bias $V_L$,
most of the incoming electron have the large energy $\omega$,
then $T_A \ll 1$ from Eq.(\ref{Andr}) and the Andreev reflection
is weak, so the N-QD-S device and the N-QD-N device have the same
turn-on/off and relaxation time. But for the small bias $V_L$, the
resonant Andreev reflection is dominant in the transport process
in the hybrid system, so that the current $J_L(t=\infty)$ of the
hybrid system is twice as that of the normal system, and their
character of $J_L(t)$ also are very different. So we will only
discuss the small pulsed bias $V_L$ case further in the following.

\begin{figure}
\includegraphics[bb=7mm 10mm 211mm 163mm, width=8.0cm,totalheight=6cm, clip=]{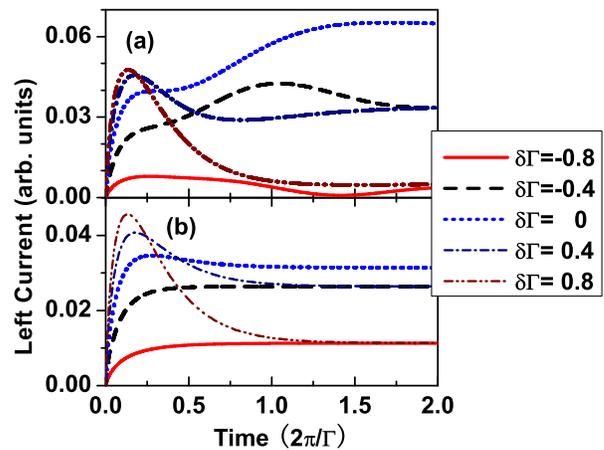}
\caption{(Color on line) The current $J_L^U(t)$ {\sl vs.} the time
$t$ in the small pulse bias $V_L=0.1$ with the different
asymmetric coupling strength $\delta\Gamma$. The panel (a) and (b)
are for the N-QD-S system and the N-QD-N system, respectively. The
other parameters are same to the Fig.1}
\end{figure}

At last, we consider the case of asymmetric barriers (i.e.,
$\delta\Gamma=\Gamma_L-\Gamma_R\neq0$) and in the small pulsed
bias $V_L$. Because in the small $V_L$ the time-dependent current
$J_L(t)$ for the upwards and downwards pulse are symmetric,
we only study the upwards case. Fig.4 plots the current $J_L^U(t)$
versus the time $t$ for the different asymmetric coupling strengths
$\delta\Gamma$, and they have the following behaviors: (i) As
$\delta\Gamma$ (i.e. $\Gamma_L$) increases, the
current $J^U_L(t)$ rises faster, i.e. the turn-on time is shorter,
because electrons with the energy in bias window can
tunnel through the left barrier more easily with the larger
$\Gamma_L$. This rising process of $J_L^U$ are nearly same for the
normal and hybrid systems. (ii) After the rise of $J_L^U$
(at $t\simeq 0.2(2\pi/\Gamma)$), Andreev reflection begins to
dominate and gives rise to different sequent relaxation processes for the
normal and hybrid systems. At $\delta\Gamma<0$, $J^U_L(t)$ of the
hybrid system humps slightly in the relaxation process, which is
obviously different from the normal system in which $J^U_L(t)$ is
monotonically relaxed into steady state. When $\delta\Gamma=0$,
$J^U_L(t)$ of the hybrid system passes a step and increases again.
The relaxation time for the hybrid system is much longer than that
of the normal system when $\delta\Gamma \leq 0$. When
$\delta\Gamma>0$, the relaxation processes of $J^U_L(t)$ are
similar for the hybrid system and the normal system. These
behaviors can be interpreted by combining the density of state
(DOS) of the QD with the Andreev reflection possibility $T_A$. In
fact, at $\delta\Gamma>0$, the DOS of the QD in the hybrid system
is similar to that of the normal system and $T_A \ll 1$, so that
the two systems have the similar turn-on/off and relaxation
characteristic. On the other hand, when $\delta\Gamma=0$ or
$\delta\Gamma<0$, the resonant or the near resonant Andreev
reflection occurs, Andreev bound states appears in the QD, and the
DOS of the QD is very different from the normal system. This makes
the relaxation processes very different for the
N-QD-S and N-QD-N systems. (iii) Although $J^U_L(t)$ for
$\delta\Gamma= +a$ and $\delta\Gamma=-a$ ($a$ is an arbitrary real
number) experience different rising and relaxation processes,
they have the same steady value at $t=\infty$. In fact, in the
steady state case and at the small bias $V_L$ limit, the
transmission possibility of the normal N-QD-N device is:
$$T(\omega)=\frac{\Gamma^2-\delta\Gamma^2}{4\omega^2+\Gamma^2},$$
and the Andreev reflection possibility of the hybrid N-QD-S device
is:\cite{Sun}
$$T_A(\omega)=\frac{(\Gamma^2-\delta\Gamma^2)^2}
{4(4\omega^2+\Gamma\delta\Gamma)^2+(\Gamma^2-\delta\Gamma^2)^2},$$
with the current expressions
$J_L=-2q\int\frac{d\omega}{2\pi}(f(\omega-V_L)-f(\omega))T(\omega)$
and
$J_L=-2q\int\frac{d\omega}{2\pi}(f(\omega-V_L)-f(\omega+V_L))T_A(\omega)$,
respectively. Here $T$ and $T_A$ are the same for $\pm \delta\Gamma$
when $\omega =0$, consequently $J_L(t=\infty)$ also are same for
$\pm \delta\Gamma$.

\section{conclusions}
In summary, we have studied the dynamic response of current to
the external upwards or downwards pulsed bias for the hybrid
N-QD-S system. In the small bias $V_L$ limit, the turn-on/off
time and the relaxation process for the upwards and the downwards
pulse bias are symmetric. Comparing wtih the normal N-QD-N system,
the Andreev reflection dominates the transport process. This makes
the turn-on/off time much longer and new steady state current
almost doubled. For the asymmetric barriers, the transport
properties of the hybrid N-QD-S system are nearly same with the
normal N-QD-N system when $\Gamma_L>\Gamma_R$. On the other hand,
while $\Gamma_L<\Gamma_R$ the current humps in the relaxation
process which reflects the properties of the superconductor.
Beyond the linear bias regime, the
rising process for upwards bias and the falling process for
downwards bias become more and more asymmetric with the increasing bias $V_L$.
The turn-on time
is faster than the turn-off time, and the current
versus the time $t$ oscillates with the frequency
$\hbar\Omega=V_L$.

$${\bf ACKNOWLEDGMENTS}$$

This work was supported by the Chinese Academy of Sciences and
NSF-China under Grant Nos. 90303016, 10474125, and 10525418. J.W. is
supported by RGC grant (HKU 7044/05P) from the government SAR of
Hong Kong and LuXin Energy Group.

\begin{appendix}
$${\bf APPENDIX}$$

In this appendix, we give the self-energy $\Sigma^{r,<}$ for
coupling to the left normal and right superconductor lead. To
consider the wide-band limit, in which the hopping elements
$t_{k,\alpha}$ is independent with the momentum $k$ and the
density of state of the leads $\rho_{L/R}^N(E)$ is energy
independent, the self-energies $\Sigma_{L,\sigma}^{r,<}$ from the
coupling to the left normal lead with the time dependent bias
potential $W_L(t)$ and in the Nambu representation are:
\begin{eqnarray}
\Sigma_{L,\sigma}^r(t',t)& = & \sum
_{k,L}t^*_{k,L}g^r_{k\sigma,L}(t',t)t_{k,L}
 = -\frac{i}{2}\Gamma_L\delta(t'-t) \label{self0} \\
 \Sigma_{L,\uparrow}^<(t',t)&=& \sum_{k,L}t^*_{k,L}g^<_{k\uparrow,L}(t',t)t_{k,L}
 \nonumber \\
 &=& i\int \frac{d \omega}{2\pi} f(\omega) \Gamma_L e^{-i\omega
 (t'-t) -
  i\int^{t'}_{t} dt_1 W_L(t_1)}   \\
 \Sigma_{L, \downarrow}^<(t',t)&=&
 \sum_{k,L}t^*_{k,L}g^<_{k\downarrow,L}(t',t)t_{k,L}
 \nonumber \\
 & =& i\int \frac{d \omega}{2\pi}
 (1-f(\omega)) \Gamma_L
e^{i\omega (t'-t) +i\int^{t'}_{t} dt_1
 W_L(t_1)}
\nonumber \\
 & =& i\int \frac{d \omega}{2\pi}
 f(\omega) \Gamma_L
e^{-i\omega (t'-t) +i\int^{t'}_{t} dt_1
 W_L(t_1)}  \label{self2}
\end{eqnarray}
Here $\Gamma_{L} =2\pi |t_{k,L}|^2 \rho_L^N$,
$g^{r,<}_{k\sigma,L}(t',t)$ is the Green function of the isolated
left lead, and $f(\omega)$ is the Fermi distribution. Notice that
the retarded self-energy $\Sigma^r_{L,\sigma}(t',t)$ is not affected
by the time dependent bias $W_L(t)$, so it is still the function of
the time difference $t'-t$. Since the time dependent bias $W(t)$ is
applied only on the left normal lead and $W_R(t)=0$, so the
self-energies for coupling to the right superconductor lead are same
with the steady state case and they can be written
as:\cite{Sun2,Sun}
\begin{eqnarray}
&&\Sigma_R^r(\omega)=-i \frac{\Gamma_R}{2}
\frac{\nu}{\sqrt{\omega^2-\Delta^2}}\left(
\begin{array}{cc}
\omega & \Delta \\
\Delta & \omega
\end{array}
\right) \\
&&\Sigma_R^<(\omega)=i\theta(\omega-\Delta)f_R(\omega)\frac{\Gamma_R}{2}
\frac{1}{\sqrt{\omega^2-\Delta^2}}\left(
\begin{array}{cc}
\omega & \Delta \\
\Delta & \omega
\end{array}
\right) \nonumber \\  \label{selfR}
\end{eqnarray}
where $\Gamma_{R} =2\pi |t_{k,R}|^2 \rho_R^N$, $\Delta$ is the
energy gap of the superconductor lead, and $\nu=1$ for
$\omega>-\Delta$ and $\nu=-1$ otherwise.
\end{appendix}


\begin{thebibliography}{00}
\bibitem{note1}
The turn on/off time describes how fast can a
device turn on/off a current, which is also named rise/fall time in the Ref.(10).
While the relaxation time was referred
to how fast can the device go to a new steady state after a bias is
abruptly switched on, it is also named saturation time in the Ref.(11).
\bibitem{foot11}
To go beyond the wide-band limit, one has to use the approach
of Ref.\onlinecite{Maciejko}.
\end{thebibliography}
\end{document}